\documentclass[aps,twocolumn,nofootinbib,showpacs,showkeys,PRD]{revtex4-1}
\usepackage{amsmath}
\usepackage{amsfonts}
\usepackage[pdftex]{graphicx}

\begin{document}
\title{Stability of the Einstein Static Universe in open cosmological models}
\author{Rosangela Canonico}
\email[E-mail: ]{canonico@sa.infn.it}
 \author{Luca Parisi}
 \email[E-mail: ]{parisi@sa.infn.it}
 \affiliation{\mbox {Dipartimento di Fisica "E.R.Caianiello", Universit\`{a} di Salerno, Via Ponte Don Melillo, I-84081 Fisciano (Sa), Italy}\\
\mbox {INFN, Sezione di Napoli, GC di Salerno, Via Ponte Don Melillo, I-84081 Fisciano (Sa), Italy}}

\begin{abstract}
The stability properties of the Einstein Static solution of General Relativity are altered when corrective terms arising from modification of the underlying gravitational theory appear in the cosmological equations. In this paper the existence and stability of static solutions are considered in the framework of two recently proposed quantum gravity models. The previously known analysis of the Einstein Static solutions in the semiclassical regime of Loop Quantum Cosmology with modifications to the gravitational sector is extended to open cosmological models where a static neutrally stable solution is found. A similar analysis is also performed in the framework of Ho\v{r}ava-Lifshitz gravity under detailed balance and projectability conditions. In the case of open cosmological models the two solutions found can be either unstable or neutrally stable according to the admitted values of the parameters.
\end{abstract}
 

\maketitle

\section{Introduction}

The Einstein Static (ES) Universe is an exact solution of Einstein's equations describing a closed Friedmann-Robertson-Walker model sourced by a perfect fluid and a cosmological constant (see, for example, \cite{Hawking:1973uf}). This solution is unstable to homogeneous perturbations as shown by Eddington \cite{Edd:1930}, furthermore it is always neutrally stable against small inhomogeneous vector and tensor perturbations and neutrally stable against adiabatic scalar density inhomogeneities with high enough sound speed \cite{Barrow:2003ni}.

In recent years there has been renewed interest in the ES Universe because of its relevance for the Emergent Universe scenario \cite{Ellis:2002we} in which the ES solution plays a crucial role, being an initial state for a past-eternal inflationary cosmological model. In the Emergent Universe scenario the horizon problem is solved before inflation begins, there is no singularity, no exotic physics is involved, and the quantum gravity regime can even be avoided. This model, relying on the choice of a particular initial state, suffers from a fine-tuning problem which is ameliorated when modifications to the cosmological equations arise but then a mechanism is needed to trigger the expanding phase of the Universe (see \cite{Lidsey:2006md}).

The existence of ES solutions along with their stability properties has been widely investigated in the framework of General Relativity for several kinds of matter ﬁelds sources (\cite{Barrow:2009sj} and references therein). ES solutions also exist in several modified gravity models \cite{Boehmer:2010xa} ranging from the Randall-Sundrum braneworld scenario \cite{Gergely:2001tn} to Gauss-Bonnet modified gravity and $f(R)$ theories \cite{Clifton:2005at}. The issue of the existence and stability of ES solutions has also been considered in the semiclassical regime of Loop Quantum Cosmology (LQC), in either the case of correction to the matter sector \cite{Mulryne:2005ef} or the case of correction to the gravitational sector \cite{Parisi:2007kv}. Recently the same issue has been also considered in the framework of Ho\v{r}ava-Lifshitz (HL) gravity \cite{Wu:2009ah} and IR modified Ho\v{r}ava gravity \cite{Boehmer:2009yz,Park:2009zra}.

When dealing with higher order modified cosmological equations, the existence of many new ES solutions is possible, whose stability properties, depending on the details of the single theory or family of theories taken into account, are substantially modified with respect to the classical ES solution of General Relativity (GR).

Often in such analysis the case of closed $(k=1)$ cosmological models is the only one considered, neglecting the intriguing possibility of static solutions in open $(k=-1)$ cosmological models. Here we point out that, due to the aforementioned corrections to the cosmological equations, open ES models may be found even in the case of a vanishing cosmological constant or when the perfect fluid has vanishing energy density. Examples are explicitly provided for the case of LQC and HL gravity.

This paper is structured as follows. In Sec. II, we consider static solutions in the framework of LQC, following and enlarging the analysis already performed in \cite{Parisi:2007kv}. It is shown that, beside the ES solution of GR, a LQC solution arises also in the case of open cosmological models; its stability is completely characterized. Following the same approach, in Sec. III we consider static cosmological solutions in the context of HL gravity with detailed balance and projectability condition. Two solutions are found along with their stability properties. In Sec. IV, some conclusions are eventually drawn. 

\section{Loop Quantum Cosmology}

In Loop Quantum Cosmology the quantization techniques borrowed by Loop Quantum Gravity, a background-independent nonperturbative quantum theory of gravity, are applied to symmetry reduced models (see \cite{Bojowald:2008zzb} and references therein).

For the sake of simplicity, in this section we consider the modified Friedmann equations arising in the semiclassical regime of LQC \cite{Ashtekar:2006es}. We consider gravitational modifications only, neglecting the inverse volume correction to the matter sector. The motivation is twofold: the analysis of this system allows a more transparent comparison with the case of GR; moreover it allows us to follow the notations introduced in \cite{Parisi:2007kv} which will also be easily used in the analysis of the HL gravity presented in the next section.

The model considered is sourced by a perfect fluid with linear equation of state $p=w \rho$ plus a cosmological constant $\Lambda$. The classical energy conservation equation still holds,
\begin{equation}
\dot{\rho}=-3\rho H (1+w), \label{EC}
\end{equation}
while the loop quantum effects lead to a modification to the classical Friedmann equation,
\begin{equation}
H^{2}=\left(\frac{\kappa }{3}\rho +\frac{\Lambda
}{3}-\frac{k}{a^{2}} \right)\left(1-\frac{\rho }{\rho
_{c}}-\frac{\Lambda }{\kappa \rho _{c}}+ \frac{3k}{\kappa \rho
_{c}a^{2}}\right)  \label{FE}
\end{equation}
and to the Raychaudhuri equation,
\begin{eqnarray}
\dot{H}&=&-\frac{\kappa }{2}\rho \left( 1+w\right) \left(
1-\frac{2\rho }{
\rho_{c}}-\frac{2\Lambda }{\kappa \rho _{c}}\right) \notag \\
&&\!\!\! + \left[1- \frac{2\rho }{\rho_{c}}-\frac{2\Lambda
}{\kappa \rho_{c}}- \frac{3\rho (1+w)}{\rho _{c}}\right]
\frac{k}{a^{2}}+\frac{6 k^{2}}{\kappa \rho _{c}a^{4}}.  \label{RE}
\end{eqnarray}
Notice that we are considering at once the $k=0$ case and the $k=\pm1$ cases \cite{Ashtekar:2006es}. Here $\kappa=8\pi G=8\pi/M_{P}^{2}$, and the critical LQC energy density is $\rho_c \approx 0.82 M_{P}^{4}\ $.

\subsection{Static solutions}

The system of Eqs.(\ref{EC})-(\ref{RE}) admits two static solutions, i.e. solutions characterized by
$\dot{a}=\dot{H}=\dot{\rho}=0$. The first solution corresponds to the standard ES Universe in GR; the second solution arises from the LQC corrective terms:
\begin{eqnarray}
&& \rho _{GR} =\frac{2\Lambda }{\kappa (1+3w)}\,,\quad
a_{GR}^{2}=\frac{2k}{\kappa \rho _{GR}(1+w)}\,,  \label{STAT1} \\
&& \rho _{LQ} =\frac{2(\Lambda-\kappa \rho _{c})}{\kappa
(1+3w)}\,,\quad a_{LQ}^{2}=\frac{2k}{\kappa \rho _{LQ}(1+w)}\,.
\label{STAT2}
\end{eqnarray}
The conditions under which these static solutions exist are summarized in Table~\ref{tab1}; they follow from $a^2>0$ and $\rho>0$. The presence of the curvature index $k$ is worth stressing, indeed the previous analysis \cite{Parisi:2007kv} can be enlarged to enclose the $k=-1$ case where the two solutions still exist.

\subsection{Stability analysis}

The stability of the solutions Eqs.(\ref{STAT1}) and (\ref{STAT2}) can be characterized using dynamical system theory and performing a linearized stability analysis. To this aim, we first have to rewrite the system of Eqs.(\ref{EC})-(\ref{RE}) in the form of a genuine dynamical system. Indeed, in these equations the three variables $a, H$ and $\rho$ appear but the actual dynamics is constrained on a two-dimensional surface described by the modified Friedmann equation. Thus, following \cite{Parisi:2007kv}, we solve Eq.(\ref{FE}) for $1/a^2$. Two solutions are found:
\begin{equation}
\frac{1}{a^{2}}=g_\pm(\rho,H) \label{2b}
\end{equation}
where
\begin{eqnarray}
g_\pm=\frac{2( \kappa\rho +\Lambda) +\kappa\rho
_{c}\left(\! 1\pm \sqrt{1-12H^{2}/\kappa\rho _{c}}\right)}{6k}.   \label{aFE1}
\end{eqnarray}
Substituting Eq.(\ref{2b}) into Eq.(\ref{RE}), we find two branches for
the time derivative of the Hubble parameter, thus the original system splits in a pair of two-dimensional nonlinear dynamical systems in the variables $\rho$ and $H$:
\begin{eqnarray}
\mbox{GR} &:&\dot{\rho}=-3H\rho \left( 1+w\right) \quad \mbox{and}
\quad \dot{H}=F_{-}(\rho ,H), \label{gr} \\ 
\mbox{LQ} &:&\dot{\rho}=-3H\rho \left( 1+w\right) \quad \mbox{and}
\quad \dot{H}=F_{+}(\rho ,H), \label{lq}
\end{eqnarray}
where
\begin{eqnarray}
F_{\pm}&=&-\frac{\kappa }{2}( 1+w)\rho \left( 1-\frac{2\rho }{
\rho_{c}}-\frac{2\Lambda }{\kappa \rho _{c}}\right)+
{6 k^{2} g_\pm^2\over \kappa\rho_c}   \notag \\
&&+  g_\pm k\left[1-{2\rho\over \rho_c }-{2\Lambda
\over\kappa\rho_c }- {3(1+w)}{\rho\over \rho_c}\right].
  \label{f1}
\end{eqnarray}
Each one of the systems  (\ref{gr}) and (\ref{lq}) admits a fixed point representing a static solution, that is,
\begin{eqnarray}
\mbox{GR} &:& H=0 \quad \mbox{and} \quad \rho_{o}=\frac{2\Lambda}{\kappa(1+3w)}, \label{FPGR}
\\
\mbox{LQ} &:& H=0 \quad \mbox{and} \quad \rho_{o}=\frac{2(\Lambda -\kappa \rho_{c})}{\kappa(1+3w)}, \label{FPLQ}
\end{eqnarray}
respectively. Substituting these values of $\rho_{o}$ in Eq.(\ref{FE}) one gets exactly the values of the constant scale factor in terms of the parameters as in Eqs.(\ref{STAT1}) and (\ref{STAT2}).

Finally, to characterize the stability of the solutions Eqs.(\ref{STAT1}) and (\ref{STAT2}) we evaluate the eigenvalues of the Jacobian matrix for the two systems Eqs.(\ref{gr}) and (\ref{lq}) at the fixed points Eqs.(\ref{FPGR}) and (\ref{FPLQ}) respectively.

\begin{figure}[h!]
\begin{center}
\includegraphics*[scale=.90]{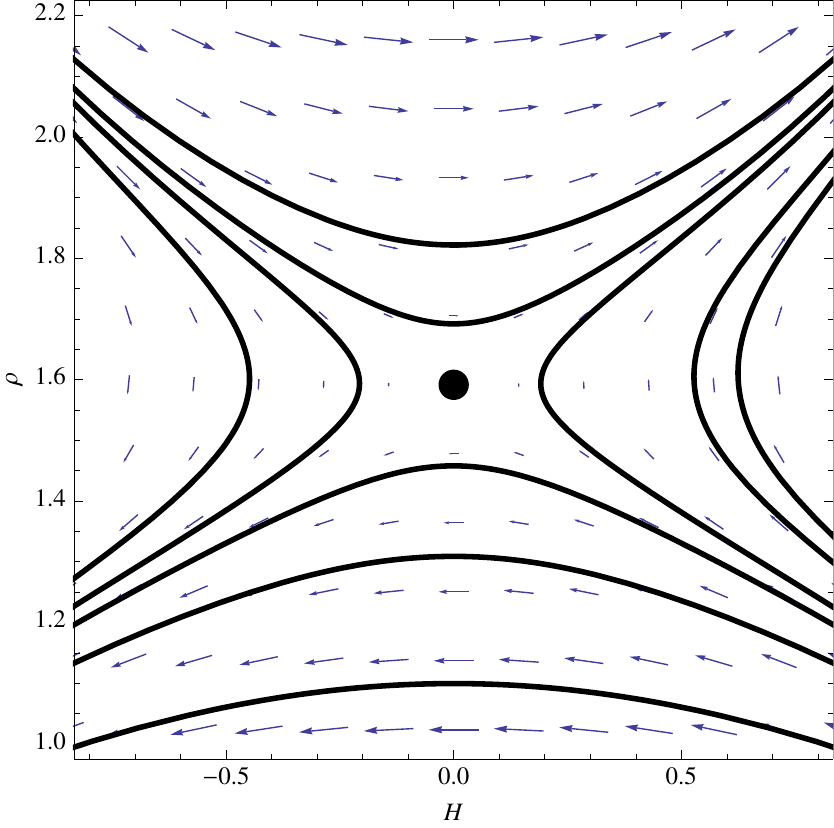}
\end{center}
\caption{Dynamical behavior of the system around the GR fixed
point for the case $k=-1$, $\Lambda<0$, $w<-1$ with
$\Lambda=-100$, $w=-2$, $\kappa=25.13274123$.} \label{1}
\end{figure}

For the system in Eq.(\ref{gr}), we recover the usual properties of the ES solution in GR. The eigenvalues of the linearized system at the fixed point are
\begin{equation}
\lambda _{GR}=\pm\sqrt{\Lambda(1+w)}.
\end{equation}
In the case of positive curvature index $k=1$, these are either real with opposite signs for $\Lambda>0$ and $w>-1/3$ - thus the fixed point is unstable (of the saddle type) - or purely imaginary for $\Lambda<0$ and $-1<w<-1/3$, so the fixed point is a center. In the case of negative spatial curvature index $k=-1$, these are again real with opposite signs for $\Lambda<0$ and $w<-1$, so the fixed point is unstable (of the saddle type). In Fig.~\ref{1} an example of the latter case is depicted.

\begin{figure}[h!]
\begin{center}
\includegraphics*[scale=.90]{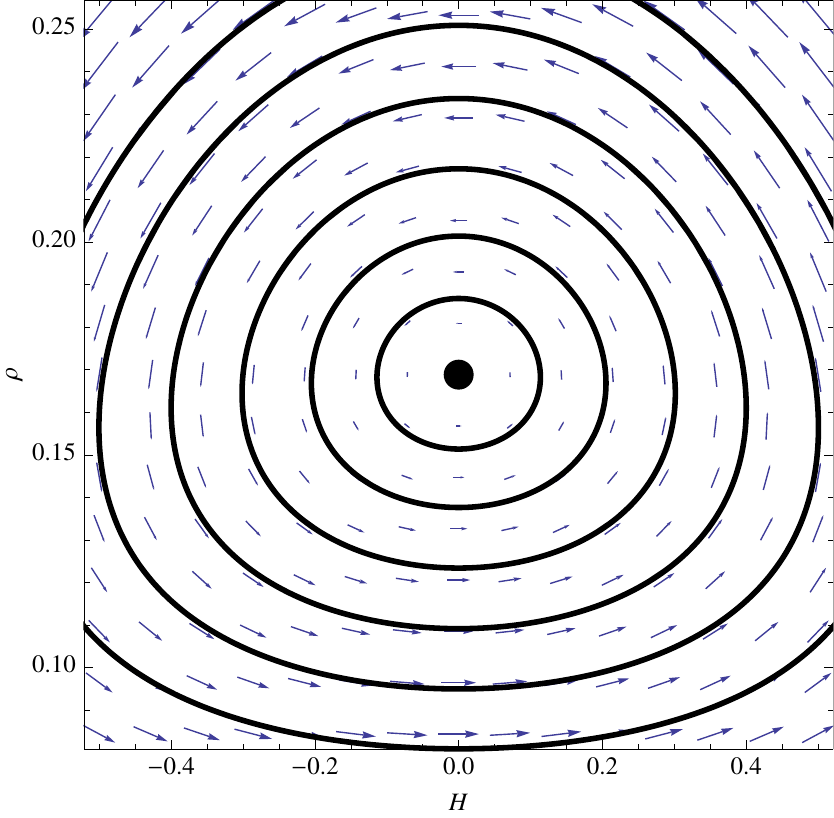}
\end{center}
\caption{Dynamical behavior of the system around the LQ fixed
point for the case $k=-1$, $\Lambda<\kappa \rho_{c}$, $w<-1$ with
$\Lambda=10$, $w=-2$, $\kappa=25.13274123$.} \label{2}
\end{figure}

For the system Eq.~(\ref{lq}) the eigenvalues at the fixed point are
\begin{equation}
\lambda _{LQ}=\pm\sqrt{(\kappa\rho_{c}-\Lambda)(1+w)}\,.
\end{equation}
In the case of positive curvature index $k=1$, the LQ fixed point is either unstable (of the saddle kind), when $\kappa\rho_{c}>\Lambda$ and $-1<w<-1/3$, or a center for the
linearized system, i.e. a neutrally stable fixed point, when
$\kappa\rho_{c}<\Lambda$ and $w>-1/3$. In the case of negative spatial curvature index $k=-1$, the eigenvalues are purely imaginary for $\kappa\rho_{c}>\Lambda$ and $w<-1$, so we have a center for the linearized system again. In the latter case the fixed point is nonhyperbolic thus the linearization theorem does not apply. Nevertheless a numerical integration of the fully nonlinear system Eq.(\ref{lq}) for initial conditions near the fixed point confirms the result of the linearized stability analysis (see Fig.~\ref{2}). It's worth stressing that in open LQC models a stable ES solution exists in the case of positive values of the cosmological constant as long as  $\Lambda< \kappa \rho_{c}$.

The results of the linearized stability analysis are summarized in Table~\ref{tab1}.

\begin{table}[h!]
\begin{center}
\begin{tabular}{|c|c|c|c|c|}
\hline
    & k  & $\Lambda$& $w$ & Stability \\  \hline
GR  & 1  & $ >0$ & $w>-1/3$    & saddle \\ \cline{3-5} 
    &   & $ <0$ & $-1<w<-1/3$ & center \\ \cline{2-5}
    & -1 & $<0$  & $w<-1$     &  saddle \\ \hline
LQ  & 1  & $<\kappa\rho_{c}$ & $-1<w<-1/3$ & center \\ \cline{3-5}
    &    & $ >\kappa\rho_{c}$& $w>-1/3$    & saddle \\  \cline{2-5}
    & -1 & $<\kappa \rho_{c}$& $w<-1$      & center \\ \hline 
\end{tabular}
\caption{Existence conditions and stability conditions for the static solutions in Eqs.(\ref{STAT1}) and (\ref{STAT2}).}
\label{tab1}
 \end{center}
\end{table}

\section{Ho\v{r}ava-Lifshitz gravity}
The Ho\v{r}ava-Lifshitz gravity \cite{Horava:2009uw} is a power-counting renormalizable theory of (3+1)-dimensional quantum gravity. In the ultraviolet limit, the theory has a Lifshitz-like anisotropic scaling between space and time characterized by the dynamical critical exponent $z=3$. In the IR limit the theory flows to the relativistic value $z=1$.

The effective speed of light $c$, the effective Newton constant $G$ and the effective cosmological constant $\Lambda$ of the low-energy theory,  emerge from the relevant deformations of the deeply nonrelativistic $z=3$ theory which dominates at short distances \cite{Horava:2009uw}:
\begin{equation}
c=\frac{\kappa^{2} \mu}{4} \sqrt{\frac{\Lambda_{W}}{1-3\lambda}}, \qquad G=\frac{\kappa^{2}}{32 \pi c}, \qquad \Lambda =\frac{3}{2} \Lambda_{W} \label{emerg}.
\end{equation}
The first of the equations in(\ref{emerg}) imposes a relation among the parameters $c$, $\Lambda_{W}$ and $\lambda$; thus, in order to have a real emergent speed of light $c$, for $\lambda>1/3$ the cosmological constant has to be negative $\Lambda_{W}$. However, after an analytic continuation of the parameters (see \cite{Lu:2009em}), a real speed of light for $\lambda>1/3$ implies a positive cosmological constant $\Lambda_{W}$. Thus, mimicking the notation introduced in \cite{Minamitsuji:2009ii}, we introduce a two-valued parameter $\epsilon=\pm 1$, in order to examine both the aforementioned cases at once.
 
The HL cosmology has been systematically studied using dynamical systems theory in \cite{Carloni:2009jc}, it has also been investigated in \cite{Wang:2009rw} using conservation laws of mechanics. Here we consider static solutions of the cosmological equations for the HL gravity when both the detailed balance condition and projectability condition hold.

First we recast the modified Friedmann equations of \cite{Lu:2009em} in a form which allows an easy comparison with the formerly considered case of LQC \footnote{According to the definitions given in Sec. II, $c=1$ and $\kappa=8 \pi G$; Eq.(\ref{FEHL}) and Eq.(\ref{REHL}) have been written accordingly.}. 

The modified Friedmann equation reads
\begin{equation}
H^{2}=\frac{2}{3\lambda-1} \left[\frac{\kappa }{3}\rho + \epsilon\left(\frac{\Lambda
}{3}-\frac{k}{a^{2}} + \frac{3 k^{2}}{4\Lambda a^{4}}\right) \right]  \label{FEHL}
\end{equation}
and the modified Raychaudhuri equation reads
\begin{eqnarray}
\dot{H}&=&  \frac{2}{3\lambda-1} \left[-\frac{\kappa }{2}\rho (1+w) + \epsilon \left( \frac{k}{a^2}-\frac{3 k^{2}}{2 \Lambda a^{4}} \right) \right].  \label{REHL}
\end{eqnarray}
The conservation equation for the energy density of the perfect fluid still holds unchanged:
\begin{equation}
\dot{\rho}=-3\rho H (1+w). \label{EC2}
\end{equation}
Besides the overall factor $ \frac{2}{3\lambda-1}$ on the right hand side of Eqs.(\ref{FEHL}) and (\ref{REHL}), the modifications to the cosmological equations of GR consist of the higher order terms $\propto k^2/\Lambda a^4$ which become dominant at short distance scales and do not affect the classical cosmological equations in the case of flat models.

\subsection{Static solutions}

It can be readily found, imposing the conditions $\dot{a}=\dot{H}=\dot{\rho}=0$, that the system of Eqs.(\ref{EC2})-(\ref{REHL}) admits the following two static solutions:
\begin{eqnarray}
&& \rho _{HL1} =0, \qquad \qquad  \qquad a_{HL1}^{2}=\frac{3k}{2 \Lambda},  \label{HLSTAT1} \\
&& \rho _{HL2} =\frac{-16 \epsilon \Lambda}{(3w-1)^{2}\kappa},   \qquad a_{HL2}^{2}=\frac{(3w-1) k }{2 \Lambda (1+w)}.
\label{HLSTAT2}
\end{eqnarray}
The conditions under which these static solutions exist are summarized in Tables \ref{tab2} and \ref{tab3}. 

The presence of the curvature index $k$ and the parameter $\epsilon$ in Eqs.(\ref{HLSTAT1}) and (\ref{HLSTAT2}) is worth being stressed; indeed the analysis presented in \cite{Wu:2009ah} can be enlarged to enclose the $k=-1$ case where new interesting possibilities arise. For instance a physically meaningful ES solution is present even in the case of vanishing energy density of the perfect fluid, i.e.  Eq.(\ref{HLSTAT1}). 

\subsection{Stability analysis}

The stability analysis can be easily performed reducing the original system to an actual two-dimensional autonomous dynamical system by making use of the Friedmann constraint. In this case the simplest and most straightforward choice is to eliminate the dependence on $\rho$ from the other equations, being Eq.(\ref{FEHL}) linear in $\rho$. This allows us to describe the dynamics with just one set of equations. Indeed, solving Eq.(\ref{FEHL}) for $\rho$,
\begin{equation}
\rho= \frac{3}{2 \kappa} \left(3 \lambda -1 \right) H^2 -\frac{ \epsilon}{\kappa} \left( \Lambda -\frac{3 k}{a^2} + -\frac{3 k^2}{4 \Lambda a^4} \right), \label{densityhl}
\end{equation}
and substituting into Eq.(\ref{REHL}) one gets a first order nonlinear differential equation,
\begin{eqnarray}
\dot{H}&=& \frac{\epsilon }{3 \lambda -1} \left[(1+w)\Lambda- \frac{(3 w+1) k}{a^2} + \frac{3 k^2 (3 w -1 ) }{4 \Lambda a^4} \right]+ \nonumber \\
& &-\frac{3}{2} \left( 1+w \right) H^2, \label{dshl1}
\end{eqnarray}
which, together with the definition of the Hubble parameter,
\begin{equation}
\dot{a}=a H \label{Hubble}
\end{equation}
provides a genuine two-dimensional autonomous dynamical system in the variables $a$ and $H$. The system admits two fixed points with energy densities as in Eqs.(\ref{HLSTAT1}) and (\ref{HLSTAT2}); thus, to characterize the stability of these solutions, we evaluate the eigenvalues of the Jacobian matrix for the system Eqs.(\ref{dshl1}) and (\ref{Hubble}) at the fixed points corresponding to Eqs.(\ref{HLSTAT1}) and (\ref{HLSTAT2}) respectively.

\begin{figure}[h!]
\begin{center}
\includegraphics*[scale=.90]{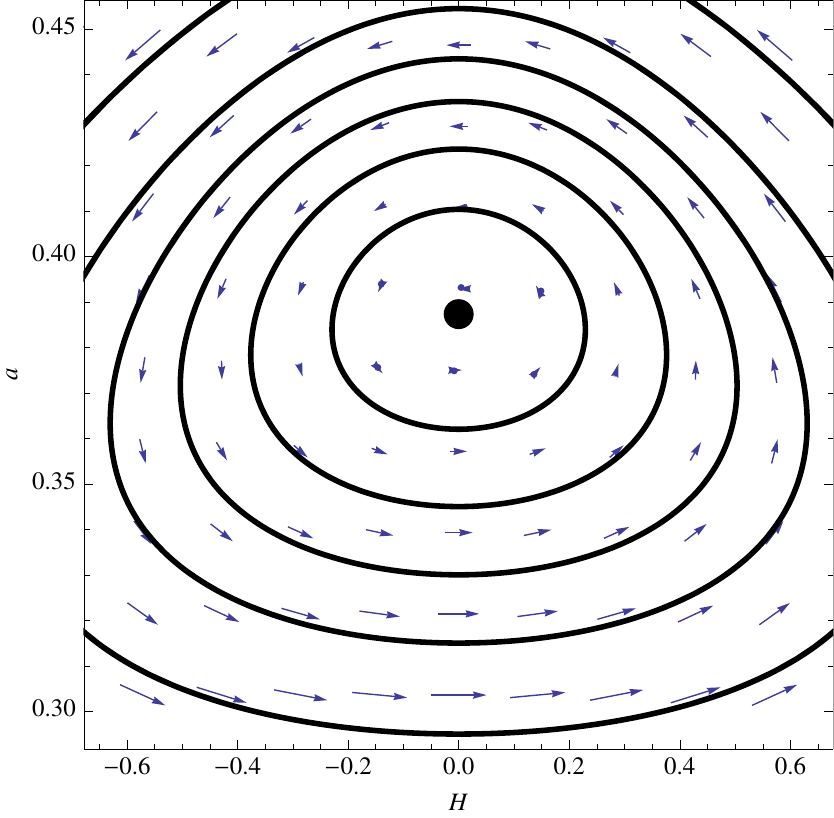}
\end{center}
\caption{Dynamical behavior of the system around the $HL1$ fixed
point for the case $k=-1$ with $\epsilon=1$,  $\lambda>1/3$, $\Lambda<0$, $w>1/3$.} \label{3}
\end{figure}

The eigenvalues at the fixed point $HL1$ read
\begin{equation}
\lambda _{HL1}=\pm \frac{2 \sqrt{6 (3 \lambda-1)\epsilon \Lambda}}{3(3 \lambda -1)}.
\end{equation}
For all the admitted values of the parameters this is a pair of purely imaginary eigenvalues thus the fixed point is a center for the linearized system. The point is nonhyperbolic so the linearized analysis may fail to be predictive at nonlinear order, nevertheless a numerical integration proves that this fixed point is actually a center (see Fig.~\ref{3}). 

The results of the stability analysis for the fixed point $HL1$ are summarized in Table~\ref{tab2}.
\begin{table}[h!]
\begin{center}
\begin{tabular}{|c|c|c|c|c|}
\hline
     $\epsilon$ & $\lambda$ & $k$  & $\Lambda$&   Stability \\  \hline
     $-1$       & $<1/3$    & $-1$  & $<0$     &   center    \\ \cline{2-4}   
                & $>1/3$    & $ 1$  & $>0$     &             \\ \cline{1-4}
     $1$        & $<1/3$    & $ 1$  & $>0$     &             \\ \cline{2-4}
                & $>1/3$    & $-1$  & $<0$     &             \\ \hline
\end{tabular}
\caption{Existence conditions and stability conditions for the static solution $HL1$.}
\label{tab2}
 \end{center}
\end{table}

\begin{figure}[h!]
\begin{center}
\includegraphics*[scale=.90]{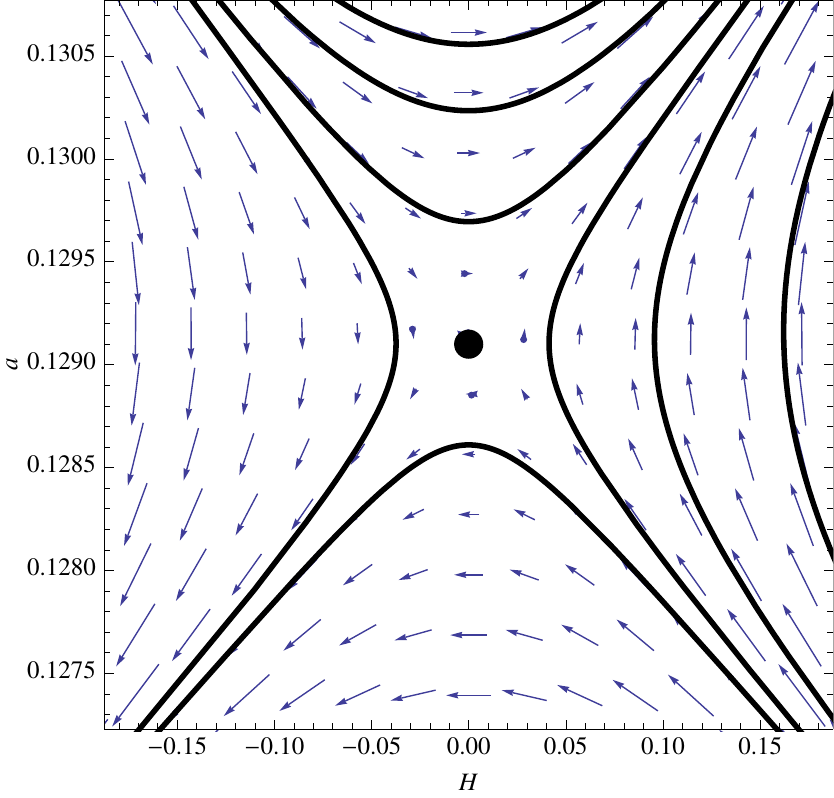}
\end{center}
\caption{Dynamical behavior of the system around the $HL2$ fixed
point for the case $k=-1$ with $\epsilon=1$,  $\lambda>0$, $\Lambda<0$, $w>1/3$.} \label{4}
\end{figure}

The eigenvalues at the fixed point $HL2$ read 
\begin{equation}
\lambda _{HL2}=\pm \frac{2 \sqrt{-2 (3w-1)(3 \lambda -1)(1+w) \epsilon\Lambda }}{(3\lambda -1)(3 w -1) }.
\end{equation}
According to the admitted values of the parameters this is either a pair of purely imaginary eigenvalues, so the fixed point is a center for the linearized system, or a pair of real eigenvalues with opposite signs, so the fixed point is unstable (of the saddle type). In particular, the solution is a center for $-1<w<1/3$ and is a saddle for $w<-1$ or $w>1/3$ (for an example of the latter case see Fig.~\ref{4}).

The results of the stability analysis for the fixed point $HL2$ are summarized in Table~\ref{tab3}.
\begin{table}[h]
\begin{center}
\begin{tabular}{|c|c|c|c|c|c|}
\hline
  $\epsilon$ & $\lambda$ & $k$  & $\Lambda$& $w$            & Stability \\  \hline 
  $-1$       & $>1/3$   & $-1$  & $>0$     &$-1<w<1/3$      & center    \\  \cline{3-6}
             &          & $ 1$  & $>0$     &$w<-1$ 			& saddle    \\  \cline{5-5}
             &          &       &          & $w>1/3$        &           \\  \hline
  $1$        & $>1/3$   & $-1$  & $<0$     &$w<-1$          & saddle    \\  \cline{5-5}
             &          &       &          & $w>1/3$        &           \\  \cline{3-6}
             &          & $ 1$  & $<0$     &$-1<w<1/3$      & centre    \\  \hline 
\end{tabular}
\caption{Existence conditions and stability conditions for the static solution $HL2$.}
\label{tab3}
 \end{center}
\end{table}

\section{Conclusions}

We have considered the existence of static solutions in the framework of two recently proposed quantum gravity models, namely, LQC and HL gravity. We have shown that the inclusion of a negative curvature index $k=-1$ enlarges the ranges of existence of the solutions affecting their stability properties thus providing new interesting results. The solutions found display stability conditions rather different from those of the corresponding solutions in closed models and from the stability properties of the standard ES solution of GR.

In the case of LQC gravitational modifications to the Friedmann equations, a negative curvature index allows a neutrally stable static solution with $\Lambda<\kappa\rho_{c}$ and $w<-1$, in contrast to the GR case. In particular the LQC static solution exists and is stable in the case of positive values of the cosmological constant as long as  $\Lambda< \kappa \rho_{c}$.

In the case of HL gravity two static solutions are found. The inclusion of the negative curvature index leads to a static solution ($HL1$) with negative cosmological constant and vanishing energy density which is neutrally stable against homogeneous perturbations. Furthermore, a negative curvature index allows a static solution ($HL2$) which can be either a saddle, for $w<-1$ and $w>1/3$, or a center for $-1<w<1/3$.

As already observed in the frameworks of different modified models \cite{Lidsey:2006md, Mulryne:2005ef, Parisi:2007kv}, the regime of infinite cycles about the center fixed points must be eventually broken in order to enter the current expanding universe phase. To this aim a further mechanism is needed, whose analysis is beyond the scope of this paper.

\begin{acknowledgements}
LP would like to thank C. G. B\"{o}hmer, F. S. N. Lobo, R. Maartens and K. Vandersloot for useful discussions. The authors would like to thank G. Vilasi for his thoughtful advice and continuous encouragement. This work is partially supported by Agenzia Spaziale Italiana (ASI), and by the Italian Ministero Istruzione Universit\`{a} e Ricerca (MIUR) through the PRIN 2008 grant. 
\end{acknowledgements}

\end{document}